\documentclass{jfm}
\usepackage{graphicx}
\usepackage{epstopdf, epsfig}
\usepackage{amsmath}
\usepackage{amsfonts}
\usepackage{amssymb}

\shorttitle{Compressibility regularizes the $\mu(I)$-rheology for granular flows}
\shortauthor{J. Heyman, R. Delannay, H. Tabuteau and A. Valance}

\title{Compressibility regularizes the $\mu(I)$-rheology for dense granular flows}

\author{J. Heyman\aff{1}
  \corresp{\email{joris.heyman@univ-rennes1.fr}},
  R. Delannay \aff{1}, H. Tabuteau \aff{1}
 \and A. Valance\aff{1}}

\affiliation{\aff{1} Institut de Physique de Rennes, UMR CNRS 6251, Université de Rennes 1, Campus de Beaulieu, Bâtiment 11A, 263 Avenue Général Leclerc, 35042 Rennes CEDEX, France}

\begin{document}

\maketitle

\begin{abstract}
The $\mu(I)$-rheology was recently proposed as a potential candidate to model the incompressible flow of frictional grains in the dense inertial regime. However, this rheology was shown to be ill-posed in the mathematical sense for a large range of parameters, notably in the low and large inertial number limits \citep{Barker2015}. In this rapid communication, we extend the stability analysis of \citet{Barker2015} to compressible flows. We show that compressibility regularizes mostly the equations, making the problem well-posed for all parameters, with the condition that sufficient dissipation be associated with volume changes. In addition to the usual Coulomb shear friction coefficient $\mu$, we introduce a bulk friction coefficient $\mu_b$, associated with volume changes and show that the problem is well-posed if $\mu_b>1-7\mu/6$. Moreover, we show that the ill-posed domain defined by \citet{Barker2015} transforms into a domain where the flow is unstable but remains well-posed when compressibility is taken into account. {These results suggest the importance of taking into account dynamic compressibility for the modelling of dense granular flows and open new perspectives to investigate the emission and propagation of acoustic waves inside these flows.}
\end{abstract}


\section{Introduction}
{The so called $\mu(I)$-rheology was recently proposed to model granular flows in the dense inertial regime \citep{GDRMidi2004,DaCruz2005}. This rheology rests on the fact that unidirectional granular shear flows are fairly well described using a single friction coefficient $\mu$---ratio of the shear stress $\tau$ to the confinement pressure $p$---that varies with an inertial (dimensionless) number $I$, defined as the ratio of a microscopic grain rearrangement time scale to a macroscopic flow time scale. This rheology may be thought as a generalization of the basic Coulomb friction model $\tau/p=\mu$, with a friction coefficient that varies according to the local shear rate and confinement pressure. 

{This simple scaling was shown to break at low inertial numbers close to the jamming limit, where non-local effects become important \citep{Kamrin2012}. On the other hand, at very high inertial numbers, granular flows enter the collisional regime and are best described by kinetic theory.} Regarding two-dimensional incompressible flows, the $\mu(I)$-rheology was also recently shown to lead to ill-posed problems for a large range of parameters, notably for high and low inertial numbers \citep{Barker2015}. For this range of parameters, infinitely small wavelengths are indeed amplified at an unbounded rate, which yields non-physical solutions.  In comparison, \citet{Schaeffer1987} and \citet{Pitman1987} showed that the constant Coulomb friction rheology is always ill-posed in the incompressible limit but that adding compressibility effects would ``greatly regularize the equations''. 

{Granular flows are indeed prone to dilate or contract in response to deformation. In the collisional limit, the volume occupied by granular flows depends directly on the granular temperature, which itself is a function of the imposed shear, thus allowing the propagation of acoustic waves. \citet{Forterre2002} showed that kinetic theory taken in the compressible limit could reproduce the instability leading to the longitudinal vortices observed in their experiments. Dense granular flows in the inertial regime might also be strongly affected by acoustic waves, resonances and instabilities due to dynamic density fluctuations as recently observed in numerical and experimental studies \citep{Melosh1979,Borzsonyi2009,Brodu2013,Trulsson2013,Krishnaraj2016}. A classic example of such instability is also found in the pulsating flows frequently observed out of silos \citep{Muite2004}.}

In this rapid communication, we extend the \citet{Barker2015} analysis and show that taking into account the weak compressibility of granular flows regularizes mostly the $\mu(I)$-rheology. We found that the problem is always well-posed providing that the energy dissipation due to volume change is sufficiently important. In the limit of incompressible flows, we recover the ill-posed criteria given by \citep{Barker2015}. When compressibility is taken into account, the flow becomes linearly unstable (so that flow structures may develop) but the problem remains well-posed.

The demonstration is organized as follows. We start by recalling the equations pertaining to the $\mu(I)$-rheology and generalize them for compressible flows. We then consider the simple case of a plane shear flow and probe the stability of the equations in the limit of perturbations of infinitely small wavelengths. We show that taking into account compressibility changes the conclusions of \citet{Barker2015} concerning the ill-posed behaviour of the $\mu(I)$ rheology. We finally validate our theoretical results by numerical resolution of the general eigenvalue problem.  

\section{Well and ill posed problems}
The distinction between well- and ill-posed (equivalently well- or ill-set) Cauchy problems is generally attributed to \citet{Hadamard}. Differential problems are termed ``well-posed'' if they yield unique solutions depending continuously on the boundary or initial conditions. These mathematical properties are extremely important for numerical modelling, insuring that numerical schemes will converge to a unique  solution, regardless of the chosen discretization.  \citet{Birkhoff1954} showed that the well-posed behaviour of a differential problem could be related to the existence of well-defined Fourier modes, for which the growth rates are bounded.

In contrast,  ill-posed problems lack continuous dependence or existence of solution. They may lead to huge and unbounded oscillations as the wavelength tends to zero, a phenomena called \textit{Hadamard instability}. For instance, the governing equations of Kelvin-Helmholtz and Rayleigh-Taylor instabilities neglecting surface tension suffer from the catastrophic short-wave instabilities of the Hadamard type \citep{Joseph1990}. Including higher-order contributions in the equations (invoking viscous terms for instance), allows generally to stabilize the instability by cutting off high wavenumbers, making the problem well-posed and suitable to numerical modelling. 

It is interesting to note that numerical schemes themselves introduce some ``viscosity'', that may regularize by chance the system and give an apparent good behaviour to numerical solutions. However, numerical ``viscosity'' obviously depends on the chosen numerical method.   Regularization has thus to be provided at a lower level, by including the missing physics in the equations. \citet{Pitman1987} showed that including compressible effects may regularize the classical Coulomb friction rheology for shear flows.

We show in the following that, under certain conditions, compressibility may also be a regularizing mechanism for the $\mu(I)$ rheology. As suggested by \citet{Birkhoff1954}, we infer the well-posed behaviour of the problem by ensuring that the growth rates of perturbations are bounded in the short-wave limit.

%
 
\section{A compressible $\mu(I)$ rheology}
Some experiments have shown that simple unidirectional shear flows the dense inertial regime present an average friction coefficient $\mu=\tau/p$ and volume fraction $\phi$ that scale with a so-called inertial number defined as 
\begin{equation}
I=\frac{d\dot{\gamma}}{\sqrt{p/\rho}},
\label{eq:I}
\end{equation} 
with $d$ the grain size, $\rho$ the material density, $\dot{\gamma}$ the shear rate, $\tau$ the shear stress and $p$ the imposed pressure \citep{GDRMidi2004,DaCruz2005,Jop2006}. Simple empirical laws were proposed such as:
\begin{eqnarray}
\mu(I)&=&\mu_s+\frac{\Delta\mu}{{I_0}/{I}+1}, \quad \phi(I)=\phi_\text{max}-\Delta_\phi I, 
\label{eq:mu(I)}
\end{eqnarray}
where $\mu_s$ is a ``static'' friction coefficient and $\phi_\text{max}$ is the maximal random packing fraction. $\Delta\mu\approx 0.3$, $I_0\approx0.3$ and $\Delta_\phi\approx 0.1$ are empirical parameters which depend on material properties \citep{GDRMidi2004,DaCruz2005,Jop2006}. 

These scaling laws were established in the hypothesis of flow incompressibility, transferring them to compressible flow requires thus some precautions. Indeed, the material density $\rho$ appears in the definition of $I$, so that the inertial number is expected to show a direct dependence upon $\phi$ through $\rho=\varrho \phi$, with $\varrho$ the grain density. This dependence may be justified physically by considering that, for equal confinement pressure, the more dilute the flow, the higher the forces transmitted through the granular skeleton. These forces scale as $d^2p/\phi$, implying a dependence of the microscopic rearrangement time scale on volume fraction. However, as the flow dilutes, rearrangement between grains are likely to take place over lengths of the order of $d/\phi$ so that both effects may eventually balance each other when the volume fraction changes. We thus chose to retain the usual incompressible definition of the inertial number taking the grain density $\varrho$ instead of the material density $\rho$. 

At this point, it is interesting to estimate the importance of the dynamical pressure developing while shearing a granular media and its possible coupling with the main flow \citep{Trulsson2013}. Based on the experimental scaling $\eqref{eq:mu(I)}$ and the definition of the inertial number \eqref{eq:I}, we may express the friction coefficient $\mu$ and the pressure at equilibrium $p_\text{eq}$ as a function of the volume fraction $\phi$:
\begin{eqnarray}
\mu(\phi)&=&\mu_s+\frac{\Delta\mu}{{I_0\Delta_\phi}/(\phi_\text{max}-\phi)+1}, \quad p_\text{eq}(\phi)=\varrho \left( \frac{d \dot{\gamma}}{\phi-\phi_\text{max}} \right)^2.
\label{eq:closure}
\end{eqnarray}
The latter expression plays the role of an equation of state for the granular flow, linking pressure and volume fraction to the kinetic properties of the medium and was assessed numerically for unidirectional shear flows \citep{Trulsson2013}. From this equation, the typical propagation speed of a pressure wave is
\begin{equation}
c=\sqrt{\frac{\partial p_\text{eq}}{\partial \rho}}=\frac{\dot{\gamma}  d \Delta_\phi}{(\phi_\text{max}-\phi)^{3/2}}.
\end{equation}
This ``dynamic'' compressibility mechanism differs significantly from acoustical wave propagation in static media in that, in the absence of shear rate, no propagation is possible ($c=0$). Taking $U=L\dot{\gamma}$ as a characteristic flow velocity ($L$ being a characteristic length scale), the Mach number of a granular flow reads
\begin{equation}
\mathcal{M}=\frac{U}{c}= \frac{L}{d}\frac{(\phi_\text{max}-\phi)^{3/2}}{\Delta_\phi}.
\end{equation} due to volume change
When approaching the random close packing fraction $\phi_\text{max}$, the wave speed diverges leading to a flow with infinitely small Mach numbers. Close to the jamming point, decoupling between flow and compressibility waves is thus likely to occur. At smaller volume fractions, Mach numbers much larger than 1 may be observed for typical parameter values observed in dense granular flows in the inertial regime (e.g., $\Delta_\phi\approx0.1$, $\phi_\text{max}\approx 0.8$, $\phi\approx 0.5-0.7$, $L/d\approx10$). Effective coupling may thus arise between the compressibility waves and the flow. In the following, we show how the $\mu(I)$ rheology may be adapted to compressible flows.

\citet{Jop2006} proposed applying the scaling laws found for unidirectional shear flows to three-dimensional flows in the incompressible limit. Given the strain rate tensor ${D}_{ij}=(\partial_i u_j+\partial_j u_i)/2$, with $u_i$ the flow velocity components, and its second invariant $\|\mathsfbi{D}\|=\sqrt{\text{tr}(\mathsfbi{D}^2)/2}$, they proposed
\begin{equation}
\frac{\boldsymbol{\tau}}{p}=\mu(I)\frac{\mathsfbi{D}}{\|\mathsfbi{D}\|},
\label{eq:incomp}
\end{equation}
with $I=2d\|\mathsfbi{D}\|/\sqrt{p/\varrho}$. This expression is similar to classical plasticity models with associated flow rule and von Mises yield criterion, but with a rate dependence of the friction coefficient \citep{Goddard2014}. \citet{Barker2015} showed that the rheology \eqref{eq:incomp} together with the incompressible equations of motion led to ill-posed problems for a large range of parameters, motivating an extension of the theory to compressible flows. 

In the compressible case, isochoric deformations (pure shear) need to be distinguished from the deformations associated with a net change in volume by splitting the strain rate tensor into deviatoric and isotropic parts, yielding
\begin{eqnarray}
D_{ij}&=&S_{ij}+\frac{1}{3}\text{tr} (\mathsfbi{D})\delta_{ij},
\end{eqnarray}
where $\delta_{ij}$ is the Kronecker symbol and $\text{tr}(\mathsfbi{S})=0$. In addition, the instantaneous pressure $p$ may now differ from the equilibrium pressure $p_\text{eq}$ so that we take the inertial number to be $I=2d\|\mathsfbi{S}\|/\sqrt{p_\text{eq}/\varrho}$. Dimensional analysis then constrains the admissible stress-strain relationship,
\begin{eqnarray}
&& \frac{\sigma_{ij}}{p_\text{eq}(\phi)} =-\delta_{ij}+ \mu(\phi) \frac{{S}_{ij}}{\|\mathsfbi{S}\|}  + \mu_b(\phi) \frac{\text{tr}(\mathsfbi{D})}{\|\mathsfbi{S}\|} \delta_{ij}, \\
&& \text{with} \quad p_\text{eq}(\phi)=\varrho \left( \frac{d \dot{\gamma}}{\phi-\phi_\text{max}} \right)^2, 
\label{eq:rheo_comp}
\end{eqnarray}
where $\mu(\phi)$ is the classical friction coefficient associated with shear and given by eq. \eqref{eq:mu(I)}, $\mu_b(\phi)$ is a bulk friction coefficient associated with non-isochoric deformations \citep{Alam1997,Nott2009,Trulsson2013} and $p_\text{eq}$ is given by the equation of state \eqref{eq:closure}. Note that Eq.~\eqref{eq:rheo_comp} simplifies to \citet{Jop2006} incompressible formulation in the case of isochoric deformations. When dilation or compression occurs, the normal stress in the medium departs from $p_\text{eq}$ by
\begin{eqnarray}
p=-\frac{\text{tr}\boldsymbol{\sigma}}{3}=p_\text{eq}(\phi) \left(1-{\mu_b(\phi)}\frac{\text{tr}(\mathsfbi{D})}{\|\mathsfbi{S}\|} \right).
\label{eq:tracesig}
\end{eqnarray}
so that $\mu_b$ directly controls the amplitude of the pressure variations in the compressible flow. In contrast to $\mu$, little is known about its value and its variations with $\phi$; as a first approximation, we take it as a constant. 

The compressible rheology \eqref{eq:rheo_comp} may be considered as general. Indeed, we show in Appendix A that, in the limit of small volume changes, it is equivalent to alternative compressible models based either on the critical state theory \citep{Jackson1983,Prakash1988} or on the dilatancy model introduced by \citet{Roux1998}.

Eq.~\eqref{eq:rheo_comp} is associated with the conservation of mass and momentum
\begin{eqnarray}
\partial_t \phi +\partial_i (\phi u_i) &=& 0 \\
{R}^2 \phi( \partial_t {u}_i +{u}_j\partial_j {u}_i ) &= &-\partial_j \sigma_{ij},
\label{eq:NSND}
\end{eqnarray}
where ${u}_i$ is a dimensionless component of the velocity vector, ${R}^2=\varrho \Phi U^2/P$ is an effective Reynolds number ($\varrho$, $\Phi$, $U$, $P$, $L$ are respectively the grain density, a unit volume fraction, a unit velocity, a unit pressure and a unit length). Since $p_\text{eq}$ is a one-to-one function of $\phi$ (Eq.~\eqref{eq:closure}), so does $I$, and we can equivalently choose the latter as the independent variable in the mass conservation equation, which transforms into
\begin{eqnarray}
\partial_t I +{u}_i\partial_i I &=& - \frac{{\phi}}{{\phi}'} \partial_i {u}_i,
\label{eq:NSND2}
\end{eqnarray}
where the primes stand for a derivative with respect to $I$. $\phi$ and $\phi'$ are then directly given by eq.~\eqref{eq:mu(I)}. This form is preferable since letting $\phi'\to 0$ allows one to probe the incompressible limit.

\section{Base flow and perturbation in the short-wave limit}
\begin{figure}
\centering \includegraphics[width=4cm]{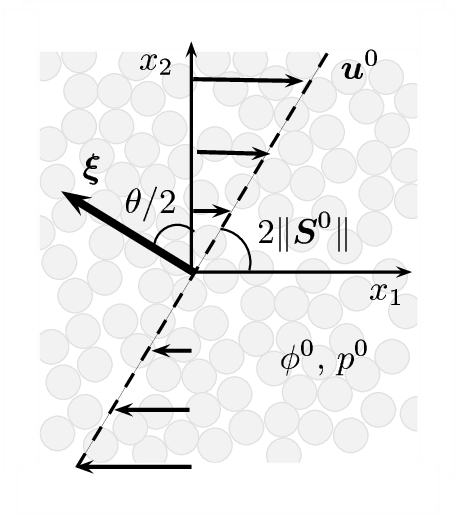}
\caption{Unidirectional shear in the plane $(x_1,x_2)$. $\boldsymbol{u}^0$ and $2\|\mathsfbi{S}^0\|$ are the flow velocity profile and shear rate. $\phi^0$ and $p_\text{eq}^0$ are the volume fraction and granular pressure. $\boldsymbol{\xi}$ is the perturbation wavevector and $\theta/2$ the inclination angle with respect to the axis $x_2$.}
\label{fig:config}
\end{figure}

We consider for the base flow a two-dimensional plane shear ($\boldsymbol{x}=(x_1,x_2)$) in absence of gravity (Fig.~\ref{fig:config}). The principal shear direction is along $x_1$ so that the velocity vector has a single non-zero component $\boldsymbol{u}=(u_1,0)$. The strain rate matrix has thus a unique non-zero component $D_{12}=\partial_2 u_1/2$. Plane shear is an isochoric deformation ($\text{tr}{\mathsfbi{D}}=0$) so that $\mathsfbi{D}=\mathsfbi{S}$. Conservation of momentum implies that the pressure and the shear stress are uniform through the flow, yielding a uniform friction coefficient $\mu$, a uniform inertial number and a uniform volume fraction. By definition of $I$, $\mathsfbi{S}$ is also uniform so that $u_1$ is a linear function of $x_2$ (Fig.~\ref{fig:config}). By choosing appropriately the scales $L$, $U$, $\Phi$ and $P$ we can always normalize the base flow so that $\|\mathsfbi{S}^0\|=1/2$, $\phi^0=1$ and $p_\text{eq}^0=1$.

The base flow solution is perturbed by short-wave normal modes of the form 
\begin{equation}
\boldsymbol{y}=\boldsymbol{y}^0+\tilde{\boldsymbol{y}}e^{\text{i} \boldsymbol{\xi \cdot  x}+\lambda t}, \quad |\boldsymbol{\xi}|\to\infty
\label{eq:pert}
\end{equation} 
where $\boldsymbol{y}=(\boldsymbol{u},I)^T$ is the vector of independent variables; $\boldsymbol{y}^0$ their values at the base state and $\tilde{\boldsymbol{y}}$ the perturbation intensity. $\boldsymbol{\xi}=(\xi_1,\xi_2)$ is the perturbation wavevector and $\lambda$ the growth rate. Note that the decomposition into normal modes in both the $x_1$ and $x_2$ direction is valid in the high-wavenumber limit because the perturbation and the base flow appear decoupled (coupling terms scale as $|\boldsymbol{\xi}|$ while leading-order terms as $|\boldsymbol{\xi}|^2$). For arbitrary wavenumbers, the coupling precludes from any analytical treatment and numerical resolution is required. Note also that we are only interested here in the early growth rate of perturbations at short time, limit at which linear analysis remains valid. Other methods are necessary to probe the asymptotic stability behaviour of the flow \citep{Alam1997}.

Inserting the perturbed variables in equations \eqref{eq:NSND}-\eqref{eq:NSND2}-\eqref{eq:closure}, and keeping only linear contributions and leading-order terms in the high-wavenumber limit (see details in Appendix B) leads to an eigenvalue problem of the form $(\mathsfbi{A}-\lambda \mathsfbi{B})\tilde{\boldsymbol{y}}=0$ with $\tilde{\boldsymbol{y}}=(\tilde{\boldsymbol{u}},\tilde{I})^T$ and $\mathsfbi{A}$ and $\mathsfbi{B}$ are $3 \times 3$ constant coefficients matrices defined by
\begin{eqnarray}
\mathsfbi{A}&=&\left(
\begin{array}{ccc}
2 \xi _1^2 \alpha +2 \mu \xi _2^2-2 \xi _1 \xi
   _2 & 2 \xi _1 \xi _2 \alpha -2
   \xi_1^2 & \text{i} \xi_2 (\beta \mu-\mu')- \text{i} \beta \xi_1 \\
 2 \xi _1 \xi _2 \alpha-2
   \xi _2^2 &
 2 \xi _2^2 \alpha+2 \mu \xi _1^2-2 \xi_1 \xi_2 & \text{i} \xi_1  (\beta \mu-\mu')- \text{i} \beta \xi_2 \\
 i \xi_1 & i \xi _2 & 0 \\
\end{array}
\right),\\
\mathsfbi{B}&=&-\left(
\begin{array}{ccc}
 {R}^2 &0&0\\
 0&{R}^2 &0\\
 0&0&\phi'
\end{array}
\right),
\end{eqnarray}
where $\mu$, $\phi$ and $I$ are taken at the base dimensionless state. The prime symbols stand for derivatives with respect to $I$ ($\phi'=\partial_I \phi$, $\mu'=\partial_I \mu$) and $\alpha={\mu_b}+2\mu/3$ and $\beta=2/I$. Non-trivial solutions to this system are obtained if $\text{det}(\mathsfbi{A}-\lambda \mathsfbi{B})=0$, which simplifies to
\begin{eqnarray}
a_3 \lambda^3+a_2 \lambda^2+a_1 \lambda+a_0 =0,
\label{eq:cubic}
\end{eqnarray}
with
\begin{eqnarray}
a_3&=& {\phi'}{R}^4, \nonumber\\
a_2&=&{2}{\phi' {R}^2}|\boldsymbol{\xi}|^2\left({\mu_b}+5\mu/3-\sin \theta \right), \nonumber\\
a_1&=& -{R}^2 |\boldsymbol{\xi}|^2 \beta \left(1-(\mu-r) \sin \theta \right ) + {\phi'}|\boldsymbol{\xi}|^4 \mu\left(4{\mu_b}+8\mu/3- 2 \sin \theta-(2{\mu_b}+\mu/3)\sin^2 \theta\right), \nonumber\\
a_0&=& |\boldsymbol{\xi}|^4\beta \left(-2r +\mu(\mu-r) \sin \theta - (\mu-2r)\sin^2 \theta  \right),
\label{eq:a}
\end{eqnarray}
where $\theta=2\tan^{-1}(\xi_1/\xi_2)$ and $r=\mu'/\beta$. Note that $\theta/2$ is the angle between the wavevector and the $x_2$ axis. To simplify notations, we dismissed the superscript over all base flow variables.


Analysis of Eq.~$\eqref{eq:cubic}$ shows that two different scaling arises for $\lambda$ at large $|\boldsymbol{\xi}|$: $\lambda\propto O(1)$ and $\lambda\propto O(|\boldsymbol{\xi}|^2)$. Both cases are considered analytically in the following.

\subsection{Case $\lambda\propto O(|\boldsymbol{\xi}|^2)$}
Writing $\lambda=\lambda' |\boldsymbol{\xi}|^2$, with $\lambda' \propto O(1)$, we retain only the $O(|\boldsymbol{\xi}|^6)$ terms in  $\eqref{eq:cubic}$ and get the quadratic equation
\begin{equation}
\lambda'^2+2a{R}^{-2} \lambda'+b{R}^{-4}=0,
\end{equation}
with 
\begin{eqnarray*}
a&=&{\mu_b}+5\mu/3-\sin\theta,\\
b&=&\mu\left(4{\mu_b}+8\mu/3- 2 \sin \theta-(2{\mu_b}+\mu/3)\sin^2 \theta\right).
\end{eqnarray*}
This yields two solutions for $\lambda$:
\begin{eqnarray}
\lambda_{1,2}&=&-\frac{|\boldsymbol{\xi}|^2 a}{{R}^{2}}\left(1\pm\sqrt{\Delta} \right), \quad \Delta=1- \frac{b}{2a^2}.
\end{eqnarray}
When $b>0$, $\Delta<1$ so that $\mbox{Re}(\lambda_{1,2})<0$ and thus the growth rate is always negative. When $b<0$, then $\Delta>1$ and one of the roots become positive with a growth rate scaling as $|\boldsymbol{\xi}|^2$, indicating ill-posed behaviour of the equations. Setting $X=1/\sin\theta$, $b>0$ is equivalent to
\begin{equation}
 4({\mu_b}+2\mu/3) X^2-2X-2{\mu_b}-\mu/3>0,\, |X|\ge 1.
\end{equation}
Since the determinant of this quadratic equation is always positive, roots are pure reals and they must lie in the interval $]-1,1[$ for $b$ to be always positive. This gives a lower bound ${\mu_b}>1-7\mu/6$ under which $b$ becomes negative and the rheology becomes ill-posed (Fig.~\ref{fig:stability}(a)). The ill-posed direction is found where $b$ takes a minimum, that is for $\theta/2=\pi/4$.

\subsection{Case $\lambda\propto O(1)$}
In this case, the growth rate $\mbox{Re}(\lambda)$ is always bounded for large $|\boldsymbol{\xi}|$, so that this particular solution does not lead to ill-posedness of the problem. To see that, as $\lambda\propto O(1)$, we retain only the $O(|\boldsymbol{\xi}|^4)$ terms and Eq.~$\eqref{eq:cubic}$ simplifies to
\begin{equation}
\lambda_3=\frac{\beta}{-\phi'b} \left(-2r +\mu(\mu-r) \sin \theta - (\mu-2r)\sin^2 \theta\right),
\end{equation}
which proves the existence of an upper bound. In other words, if $\lambda_3$ becomes positive, the flow becomes unstable in the high-wavenumber limit but the problem remains well-posed. For $b>0$ ($b<0$ was proven to generate ill-posedness) and $\phi'<0$, the sign of $\lambda_3$ is determined by the sign of the nominator. The latter is negative if
\begin{equation}
-2r X^2 +\mu(\mu-r)X-(\mu-2r)<0, |X|>1.
\end{equation}
This condition is verified if (i) the determinant $\Delta=\mu^2\left(\mu-r \right)^2-8r\left(\mu -r\right)$ is negative or (ii) the determinant is positive, but the real roots lie inside the interval $[-1,1]$, that is if $\sqrt{\Delta}<2r-\mu(\mu-r)$. Thus, $\lambda_3$ is negative if
\begin{equation}
\mu^2\left(\mu-r\right)^2-8r\left(\mu -r\right)<\text{max}\left(0,2r-\mu\left(\mu-r\right)\right)^2.
\end{equation} 
It is easy to show that $0<r<\Delta\mu/8$ and thus, for usual rheology parameters ($\Delta\mu\approx0.26$, $\mu_s\approx 0.38$), $2r-\mu\left(\mu-r\right)$ is always negative and the stability condition simplifies to
\begin{equation}
\mu^2\left(\mu-r\right)^2-8r\left(\mu -r\right)<0,
\label{eq:stab_comp}
\end{equation} 
The stability domain is shown in Fig.~\ref{fig:stability}(b) in the plane $(I/I_0,\Delta \mu)$.
\begin{figure}
\centering \includegraphics[width=\linewidth]{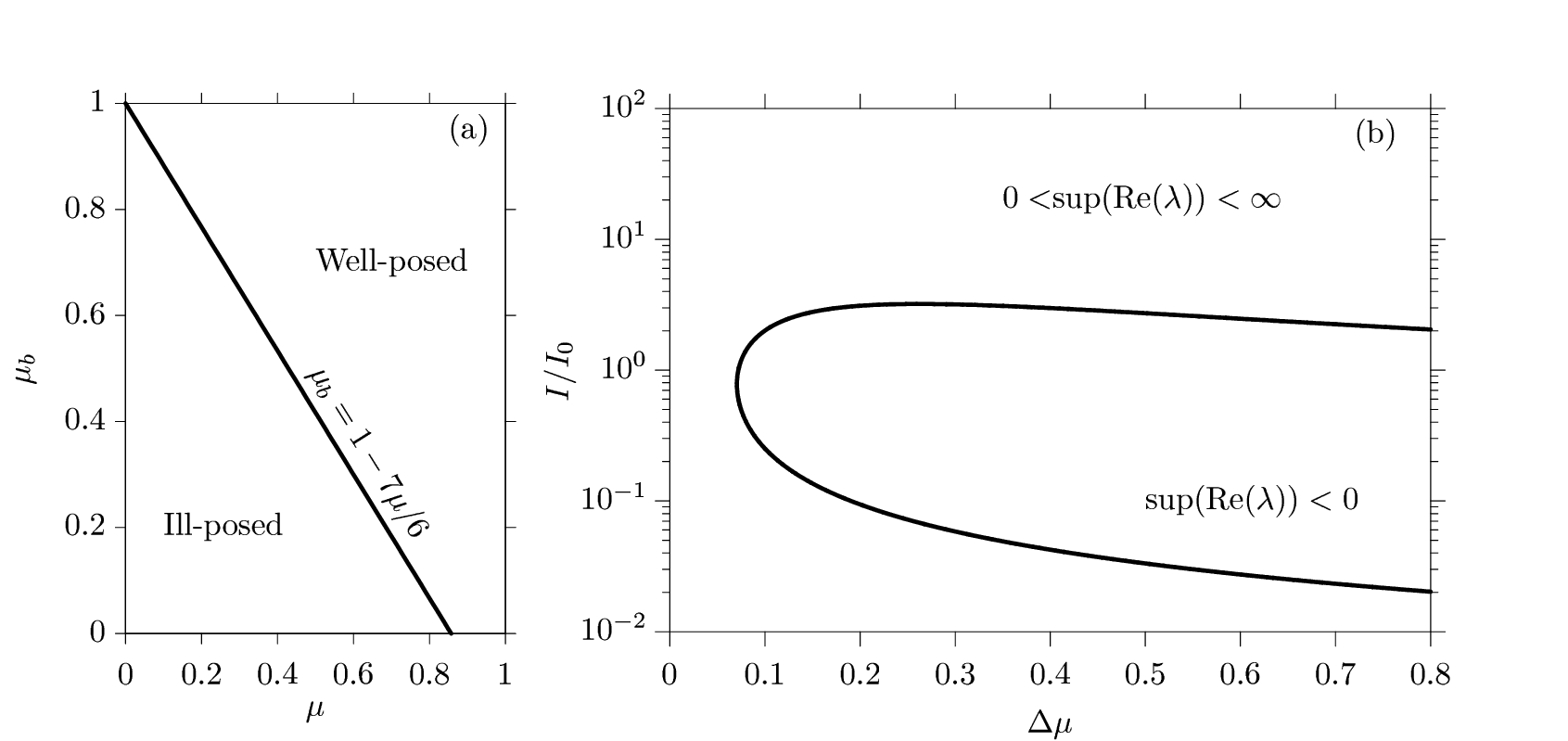}
\caption{(a) Well-posed and ill-posed domains in the $\mu -{\mu_b}$ plane.  (b) Domain of linear stability for short wavelength in the compressible $\mu(I)$-rheology as given by the inequality \eqref{eq:stab_comp}. $\mu'$ is given by the empirical scaling \eqref{eq:mu(I)} so that the inequality can be uniquely represented by the black curve in the $I/I_0 - \Delta \mu$ plane.}
\label{fig:stability}
\end{figure}

\subsection{Summary and discussion}
To summarize, the compressible $\mu(I)$-rheology leads to plane shear flows that are unstable outside of the domain defined by Eq.~$\eqref{eq:stab_comp}$ (Fig.~\ref{fig:stability}). However, if compressibility is associated with sufficient dissipation (i.e., ${\mu_b}>1-7\mu/6$) the growth rate of the unstable modes remains bounded in the high wavenumber limit, and the rheology remains well-posed at any inertial number.

It is instructive to note that the stability domain in the high wavenumber limits of compressible flows defined by \eqref{eq:stab_comp} is independent of flow compressibility $\phi'$ and equivalent to the well-posed domain defined by \citet{Barker2015} in the incompressible limit. Another way to see this is by letting $\phi'\to 0$ in Eq.\eqref{eq:a}. The growth rate of the unique eigenvalue $\lambda=-{a_0}/{a_1}$ is now scaling as $|\boldsymbol{\xi}|^2$ so that, when $\lambda$ becomes positive outside of the domain defined by \eqref{eq:stab_comp}, the incompressible rheology becomes ill-posed. In the compressible case however, the growth rate of the positive eigenvalue is bounded since $\lambda_3 \propto O(1)$.

Another interesting limit is obtained putting $\mu'=0$, which yields the classical Coulomb constant friction model. In this case, it is immediate to see that $\eqref{eq:stab_comp}$ is never verified for $\mu>0$ so that compressible shear flows described by a simple Coulomb friction are always unstable but remain well-posed since the condition ${\mu_b}>1-7\mu/6$ is not affected by the value of $\mu'$.

Let us infer afterwards the impact of embracing the volume fraction variations in the definition of the inertial number through the material density. Taking $I=2d\|\mathsfbi{S}\|/\sqrt{p_\text{eq}/(\varrho\phi)}$ has the sole effect of changing the coefficient $\beta$ to $2/I-\phi'$, making thus the stability domain $\eqref{eq:stab_comp}$ dependent of the particular value chosen for compressibility. In contrast, the well-posedness criteria is not affected by this other possible definition of the inertial number so that it can be considered as general.

A last remark can also to be made on the choice, in the expression of the inertial number, of the equilibrium pressure rather than the instantaneous pressure \eqref{eq:pressure}. As long as a linear analysis is concerned, choosing the latter has no impact on the stability and well-posed criteria derived above, since the terms introduced in the expansion are all of higher order.

\section{Numerical solution at arbitrary wavenumbers}
To test the validity of the normal mode decomposition in the short wave limit and the obtained criteria for $\mu_b$, we solve numerically the stability problem at arbitrary wavenumbers. To that end, we consider the case of a granular media sheared between two plates located at $x_2= \pm 1/2$ (in dimensionless units). As a coupling arises between the base flow and the perturbation in the sheared direction, we look for generic perturbations of the base flow of the form
\begin{equation}
\boldsymbol{y}=\boldsymbol{y}^0(x_2)+\tilde{\boldsymbol{y}}(x_2)e^{\text{i} \xi_1 x_1+\lambda t}.
\label{eq:normod2}
\end{equation}
Inserting the solution \eqref{eq:normod2} into mass and momentum conservation equations and keeping only linear contributions leads to a system of second-order ordinary differential equations in the variable $\tilde{\boldsymbol{y}}=(\tilde{\boldsymbol{u}},\tilde{I})^T$, of the form
\begin{eqnarray}
\left\lbrace \begin{array}{l}
\mathsfbi{C}_2\frac{\text{d}^2 \tilde{\boldsymbol{y}}}{\text{d} x_2^2}+
\mathsfbi{C}_1\frac{\text{d} \tilde{\boldsymbol{y}}}{\text{d} x_2}+
\mathsfbi{C}_0(x_2)\tilde{\boldsymbol{y}}=
-\lambda \mathsfbi{C}_\lambda \tilde{\boldsymbol{y}}, \\
\tilde{\boldsymbol{u}}(\pm 1/2)=0
\end{array}\right.
\label{eq:ODE}
\end{eqnarray}
where the $\mathsfbi{C}_i$ matrices are given by
\begin{eqnarray}
&&\mathsfbi{C}_2=\left(\begin{array}{ccc}
-2\mu & 0 & 0\\2 & -2\alpha & 0\\0&0&0
\end{array}\right),\quad
\mathsfbi{C}_1=\left(\begin{array}{ccc}
-2\text{i}\xi_1&-2\text{i}\alpha\xi_1 & \mu-\mu'\\-2\text{i}\alpha\xi_1 & 2\text{i}\xi_1 & -\beta\\0&1&0
\end{array}\right), \\
&&\mathsfbi{C}_0=\left(\begin{array}{ccc}
2\alpha\xi_1^2+\text{i}\mathcal{R}^2\xi_1 u_1^0&\mathcal{R}^2-2\xi_1^2 & -\text{i}\beta\xi_1\\0 & 2\mu\xi_1^2+\text{i}\mathcal{R}^2\xi_1 u_1^0 & \text{i}\xi_1(\mu\beta-\mu')\\\text{i}\xi_1&0&\text{i}\phi'\xi_1 u_1^0
\end{array}\right),\quad
\mathsfbi{C}_\lambda=\left(\begin{array}{ccc}
\mathcal{R}^2&0&0\\0 & \mathcal{R}^2 &0\\0&0&\phi'
\end{array}\right), \nonumber
\end{eqnarray}
with $u_1^0=x_2$, the base flow velocity. Coupling between base flow and perturbation clearly appears in the term $C_0(x_2)$, precluding a simple analytical treatment. The system \eqref{eq:ODE} is discretized in the direction $x_2$ and solved numerically via a Chebychev collocation method. The perturbed velocities and the momentum equations are approximated on $N$ Gauss-Lobatto collocation points while $\tilde{I}$ and the continuity equation are solved on a staggered grid made of $N-1$ Gauss collocation points \citep{Khorrami1991}. No-slip and no-penetration boundary conditions are imposed  at $x_2 \pm 1/2$ for the perturbed velocity. $\tilde{I}$ being computed on the staggered grid, its value is free to adapt at the boundary. The discretized system and the boundary conditions, reduce to an eigenvalue problem of dimension $3(N-1)$:
\begin{equation}
(\boldsymbol{A}'-\lambda \mathsfbi{B}')X=0,
\end{equation}
$X$ being the discretized version of $\tilde{\boldsymbol{y}}$ which is solved with the QR algorithm. Grid convergence of solutions is reached for $N>3\xi_1$, keeping a minimum of $20$ grid points. Following the value of $\text{sup}({\text{Re}(\lambda)})$ while varying $\xi_1$ allows us to probe the stability of the coupled system depending on its parameters (Fig.~\ref{fig:numer}a-b). The ill-posedness of the system clearly appears for $\xi_1 > 10$, where the growth rate of perturbation dramatically increase as $\xi_1^2$. Increasing the bulk viscosity has the effect of regularizing the system against short waves. It is interesting to note that in this case, while remaining bounded, growth rates may not cancel at infinitely large wavenumbers (Fig~\ref{fig:numer}b). In contrast to the low bulk viscosity case however, the maximum growth rate is now located at a finite wavenumber, giving an upper bound for the necessary grid size in order to capture the evolution of the most unstable mode in numerical simulations.

To check the validity of the condition on $\mu_b$, we compute $r_{\xi_1}=\text{sup(Re(}\lambda$)) for $\xi_1=30$, $40$ and $50$. If $r_{50}-r_{40}>r_{40}-r_{30}$ (increasing growth rate of $\text{sup(Re(}\lambda$), the system is said to be ill-posed. Inversion of the inequality provides the condition for well-posedeness. We explored the domain of $\mu$ and $\mu_b$ values and found that the condition $\mu_b\geq 1-7\mu/6$ for well-posedeness is verified almost everywhere (Fig.~\ref{fig:numer}c).

It is interesting to note that compressible shear flows appear stable in the high-wavenumber limit even outside of the condition for stability defined by Eq.~\eqref{eq:stab_comp} (right part of Fig.~\ref{fig:numer}c). This may be caused by the role played by the boundaries in filtering the admissible wavelengths. Indeed, the high-wavenumber limit considered above do not carry information on the presence of boundary conditions, that may have a stabilizing role on the flow. This said, the ill-posed criteria $\mu_b<1-7\mu/6$ is not modified by such finite size effect and can thus be considered as general.

\begin{figure}
\includegraphics[width=\linewidth]{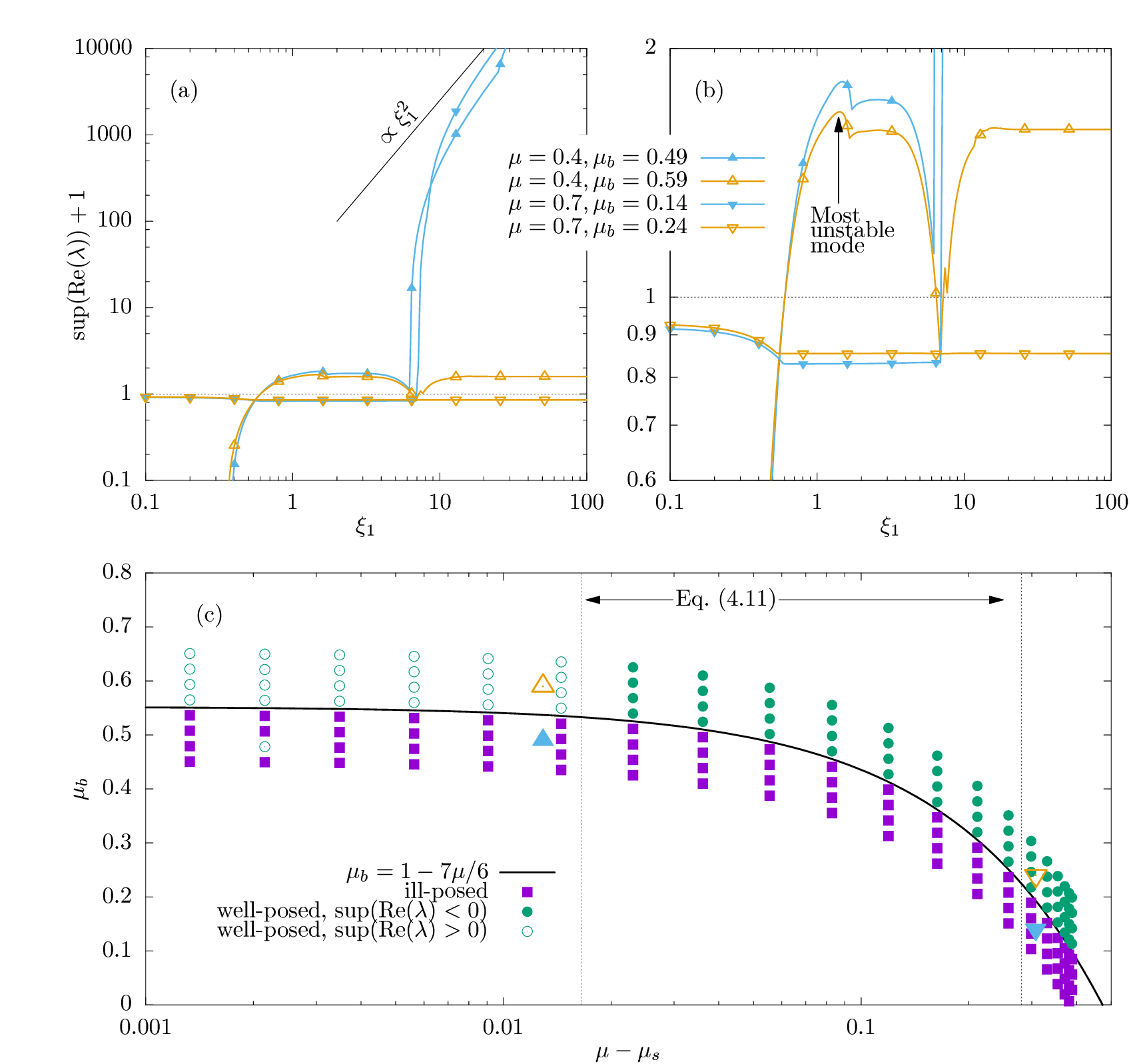}
\label{fig:numer}
\caption{(a) Maximum real eigenvalue of the system \eqref{eq:ODE} depending on the wavenumber (constant parameters are $n=2$, $R=0.1$, $\phi'=-0.5$, $\mu_s=\tan(21^\circ)$, $\Delta\mu=0.4$, $I_0=0.3$). Black line indicates the $\xi_1^2$ scaling in the ill-posed problem, while the dashed line indicates marginal stability. (b) Close up of (a) showing that, in the compressible rheology, the most unstable mode is found at $\xi_1=1.4$ for $\mu=0.4$ and $\mu_b=0.59$. (c) Numerical simulations of various $(\mu_b,\mu)$ pairs. Well-posedness is assessed by comparing $\text{sup}(\lambda_{\xi_1})$ at 3 large wavenumbers: $\xi_1=30,40$ and 50. For $\text{sup}(\lambda_{50})<0$ or $\text{sup}(\lambda_{40})-\text{sup}(\lambda_{30})>\text{sup}(\lambda_{50})-\text{sup}(\lambda_{40})$ (bounded growth rate), the system is assumed well-posed, and ill-posed otherwise. The four triangles in figure 3b (up and down) indicate where stands, in the parameter space, the computations presented in the left figure. }
\end{figure}

\section{Conclusion}
In this article we showed that introducing compressibility effects generally regularizes the $\mu(I)$-rheology proposed for dense inertial granular flows.  Volume changes need to be associated with a sufficient dissipation to make the problem well-posed at all inertial numbers. This condition is expressed in terms of an inequality between ${\mu_b}$, the bulk friction coefficient associated with volume changes and $\mu$ the friction coefficient associated with shear, which reads ${\mu_b}>1-7\mu/6$.

While being generally well-posed, the compressible $\mu(I)$-rheology was also shown to be unstable for some flow conditions, both at large and small inertial number, calling for a deeper analysis of the unstable modes appearing in compressible flows. Indeed, secondary flows and resonances of dense inertial granular flows may originate from such instabilities \citep{Borzsonyi2009,Trulsson2013,Brodu2014,Krishnaraj2016}. Although the asymptotic analysis at large wavenumbers provides precious information on the good behaviour of the problem, it is not sufficient to probe the general linear stability picture of the flow, neither its asymptotic stability at long times. To go further, the full dispersion relation needs to be solved numerically to find the most unstable wavelength which will predominate in the system.

Finally, it is legitimate to wonder if any hypothetical variations of $\mu_b$ with $I$ or other flow variables would modify significantly the preceding results. In the case of a pure shear base flow, the contributions arising from the variations of $\mu_b$ would not appear in the linearized equations so that our conclusions are still valid. In any case, additional experimental and numerical work is needed to quantify the bulk friction associated with volume changes in dense granular flows. 

In addition to provide a possible way to regularize the short-wave instabilities appearing in numerical simulations of the incompressible $\mu(I)$ rheology, this study more generally shows that the generation and propagation of acoustic waves inside granular flows are likely to play a crucial role in their rheology. Depending on their geometry and their elastic properties, boundaries may also absorb or restitute part of the acoustic energy to the flow and produce apparent non-local effects. In this aspect, the acoustic probing and characterization of flowing granular media form a rather new and promising perspective to improve our knowledge of dense granular flows. 

\section{Acknowledgement}
We acknowledge the fact that, by the time our paper was considered for publication in Journal of Fluid Mechanics, a similar study had been published in another journal \citep{Barker2017}. In this study, compressibility is introduced based on the critical state concept borrowed from soil mechanics while we preferred a fluid mechanics-like formulation based on a second viscosity (both formulations are shown to be equivalent in Appendix A). Although the methods diverge, both studies points to the same conclusion---that compressibility generally regularizes the $\mu(I)$ rheology, providing that certain criteria on the parameters are respected.

\appendix
\section{Comparison of compressibility models}
\subsection{Critical state theory}
In plasticity, a material is assumed to flow on a yield surface taking the general form  
\begin{eqnarray}
\|\boldsymbol{\tau}\|=\mu p_\text{eq}(\phi) \mathcal{F}(p/p_\text{eq}(\phi)),
\label{eq:yieldsurf}
\end{eqnarray}
where $p$ is the total normal stress, $\|\boldsymbol{\tau}\|$ is the shear stress, $\phi$ is the volumic fraction, $p_\text{eq}(\phi)$ the pressure at the critical state (when the material deforms without volume change) at the volume fraction $\phi$ (given by an empirical equation of state), $\mu$ is the friction coefficient and $\mathcal{F}$ a function controlling the curvature of the yield surface close to the critical state. \citet{Prakash1988} proposed
\begin{eqnarray}
\mathcal{F}(x)=mx-(m-1)x^{m/(m-1)},
\label{eq:F}
\end{eqnarray}
where $1<m<\mu$ is an empirical constant. Postulating the existence of an associated flow rule (co-axiality of stain rate and stress tensors), the rate of volume change is linked to the gradient of the yield surface by 
\begin{eqnarray}
\nabla \cdot \boldsymbol{v} \equiv \text{tr}(\mathsfbi{D})= \|\mathsfbi{S}\| \frac{\partial \|\boldsymbol{\tau}\|}{\partial p}. 
\label{eq:flowrule}
\end{eqnarray}
In addition, it is required that the derivative of $\mathcal{F}$ is a monotonic function so that $p$ can be uniquely determined from Eqs.~\eqref{eq:flowrule},\eqref{eq:F} and \eqref{eq:yieldsurf}:
\begin{eqnarray}
p=p_\text{eq}(\phi)\left(1-  \frac{1}{m \mu} \frac{\text{tr}(\mathsfbi{D}) }{\|\mathsfbi{S}\|}\right)^{m-1}.
\label{eq:pressure}
\end{eqnarray}
The stress-strain relationship then reads
\begin{eqnarray}
\frac{\boldsymbol{\sigma}}{p_\text{eq}(\phi)}=-\boldsymbol{\delta} \left(1- \frac{1}{m \mu} \frac{\text{tr}(\mathsfbi{D}) }{\|\mathsfbi{S}\|}\right)^{m-1} +\mu \mathcal{F} \frac{\mathsfbi{S}}{\|\mathsfbi{S}\|} ,
\end{eqnarray}
with $\boldsymbol{\delta}$ the identity tensor \citep{Krishnaraj2016}.
Linearizing for small volume changes gives
 \begin{eqnarray}
\frac{\boldsymbol{\sigma}}{p_\text{eq}(\phi)}=-\boldsymbol{\delta} +\mu \frac{\mathsfbi{S}}{\|\mathsfbi{S}\|}+ \mu_b  \frac{\text{tr}(\mathsfbi{D}) }{\|\mathsfbi{S}\|} \boldsymbol{\delta},
\label{eq:stress}
\end{eqnarray}
with $\mu_b=(m-1)/(m \mu)$, the equivalent of a bulk friction.  Eq.~\eqref{eq:stress} is equivalent to Eq.~\eqref{eq:rheo_comp}.

\subsection{Dilatancy model}
Based on geometrical considerations, \citet{Roux1998} proposed a description of the deformation of granular flows through a dilatancy angle $\psi$
\begin{equation}
-\frac{1}{\phi}\frac{\text{d} \phi}{\text{d} t}=\|\mathsfbi{S}\| \tan \psi \equiv \text{tr}(\mathsfbi{D}),
\label{eq:roux}
\end{equation}
where $\tan \psi$ is taken as a function of $\phi-\phi_\text{eq}$ \citep{pailha2009}
\begin{equation}
\tan \psi=K(\phi-\phi_\text{eq}).
\end{equation}
The dilatancy also modifies the shear stress 
\begin{equation}
{\tau}=\tau_\text{eq}+\tan \psi p_\text{eq}(\phi_\text{eq}),
\label{eq:stressRoux}
\end{equation}
with $p_\text{eq}$, the pressure at equilibrium. It is possible to interpret this model as a compressible rheology of the type \eqref{eq:rheo_comp} by equating $\text{tr}(\mathsfbi{D})$ in Eq.~\eqref{eq:tracesig} and Eq.\ref{eq:roux} to yield
\begin{equation}
\tan \psi=\frac{1}{\mu_b}\left(1-\frac{p}{p_\text{eq}}\right),
\end{equation} 
which, upon linearization around $\phi=\phi_\text{eq}$ gives
\begin{equation}
\tan \psi\approx\frac{1}{\mu_b}\frac{\phi - \phi_\text{eq}}{p_\text{eq}} \left. \frac{\partial p_\text{eq}}{\partial\phi}\right|_{\phi_\text{eq}}.
\end{equation} 
By analogy with Eq.\eqref{eq:stressRoux}, we can deduce the value of the bulk friction in the model of \citet{Roux1998}
\begin{equation}
\mu_b\approx \frac{1}{K p_\text{eq}}\left. \frac{\partial p_\text{eq}}{\partial\phi}\right|_{\phi_\text{eq}}
\end{equation}

\begin{eqnarray}
\text{tr}(\mathsfbi{D}) &=& \|\mathsfbi{S}\|  \tan \psi, \quad \text{with} \quad \tan \psi = K (\phi-\phi_\text{eq}),
\end{eqnarray}
where $\psi$ is an angle of dilatancy and $\phi_\text{eq}$ is the volumic fraction at the critical state. This formulation shares similarities with Eqs.~\eqref{eq:flowrule} and \eqref{eq:yieldsurf}, but differs in that $p$ do not appear explicitly in the equations. Alternatively, \citet{Bouchut2016} proposed
\begin{eqnarray}
\tan \psi = K_p(p_\text{eq}(\phi)-p),
\end{eqnarray}
yielding a stress-strain rate relationship similar to Eq.~\eqref{eq:stress} with $\mu_b=1/K_p$ and $\mu=(\tan \delta + \tan \psi)$.

\section{Derivation of the eigenvalue problem}
The perturbed deformation rates read
\begin{eqnarray}
\tilde{\mathsfbi{D}}&=&\text{i} \left(\begin{array}{cc}
\xi_1 \tilde{u}_1 & \frac{1}{2} (\xi_2 \tilde{u}_1 + \xi_1 \tilde{u}_2) \\
\frac{1}{2} (\xi_2 \tilde{u}_1+ \xi_1 \tilde{u}_2) & \xi_2 \tilde{u}_2 \\
\end{array}\right),  \\
 \tilde{\mathsfbi{S}}&=&\text{i} \left(\begin{array}{cc}
\frac{1}{3}(2\xi_1 \tilde{u}_1-\xi_2 \tilde{u}_2) & \frac{1}{2} (\xi_2 \tilde{u}_1 + \xi_1 \tilde{u}_2) \\
\frac{1}{2} (\xi_2 \tilde{u}_1 + \xi_1 \tilde{u}_2) & \frac{1}{3}(\xi_2 \tilde{u}_2-2\xi_1 \tilde{u}_1) \\
\end{array}\right).
\end{eqnarray}

Since $\mu$ and $\phi$ are considered sole functions of $I$, we write $\tilde{\mu}=\mu'\tilde{I}$ and $\tilde{\phi}=\phi' \tilde{I}$, where the primes stand for a derivative with respect to $I^0$. To first order, we find $\|\tilde{\mathsfbi{S}}\|=2\|\mathsfbi{S}^0\| |\tilde{S}_{12}|$ and the definition of $I$ gives
\begin{equation}
\frac{\tilde{I}}{I^0}=\frac{\|\tilde{\mathsfbi{S}}\|}{\|\mathsfbi{S}^0\|}-\frac{\tilde{p}_\text{eq}}{2p_\text{eq}^0},
\end{equation}
which allows us to eliminate $\tilde{p}_\text{eq}$ in the equations.
Given the rheological law $\eqref{eq:rheo_comp}$, the stresses are, to first order,
\begin{eqnarray}
\tilde{\sigma}_{ij}&=&\left\lbrace
 \begin{array}{l}
-\tilde{p}_\text{eq}+{p_\text{eq}^0}(\mu^0 \tilde{S}_{ij}+{\mu_b} \text{tr}(\tilde{\mathsfbi{D}}))/{\|\mathsfbi{S}^0\|} \quad \text{for} \quad i=j,\\
\tilde{\sigma}_{ij}=\mu^0 \tilde{p}_\text{eq} +\mu' p_\text{eq}^0 \tilde{I}  \quad \text{for} \quad i \neq j,\\
\end{array}
 \right.
\end{eqnarray}

\begin{table}
\centering\begin{tabular}{r|cc}
Equations & $\tilde{\boldsymbol{u}}$ & $\tilde{I}$ \\
\hline
Continuity & $O(|\boldsymbol{\xi}|)$ &  $O(|\boldsymbol{\xi}|)$  \\
Momentum & $O(|\boldsymbol{\xi}|^2)$ &  $O(|\boldsymbol{\xi}|)$ 
\end{tabular}
\caption{Order of the leading terms in equations \eqref{eq:NSND} depending on the variable and the equation for $|\boldsymbol{\xi}|\to \infty$. }
\label{tab:1}
\end{table}

Normalizing with respect to the base flow ($\|\mathsfbi{S}^0\|=1/2$, $\phi^0=1$ and $p_\text{eq}^0=1$), inserting the perturbed variables and stresses into \eqref{eq:NSND} and keeping only linear contributions and leading-order terms in the high-wavenumber limit (according to Table~\ref{tab:1}) leads to an eigenvalue problem of the form $(\mathsfbi{A}-\lambda \mathsfbi{B})\tilde{\boldsymbol{y}}=0$ with $\tilde{\boldsymbol{y}}=(\tilde{\boldsymbol{u}},\tilde{I})^T$ and
\begin{eqnarray}
\mathsfbi{A}&=&\left(
\begin{array}{ccc}
2 \xi _1^2 \alpha +2 \mu \xi _2^2-2 \xi _1 \xi
   _2 & 2 \xi _1 \xi _2 \alpha -2
   \xi_1^2 & \text{i} \xi_2 (\beta \mu-\mu')- \text{i} \beta \xi_1 \\
 2 \xi _1 \xi _2 \alpha-2
   \xi _2^2 &
 2 \xi _2^2 \alpha+2 \mu \xi _1^2-2 \xi_1 \xi_2 & \text{i} \xi_1  (\beta \mu-\mu')- \text{i} \beta \xi_2 \\
 i \xi_1 & i \xi _2 & 0 \\
\end{array}
\right),\\
\mathsfbi{B}&=&-\left(
\begin{array}{ccc}
 {R}^2 &0&0\\
 0&{R}^2 &0\\
 0&0&\phi'
\end{array}
\right),
\end{eqnarray}
where we used the simplified notation $\mu\equiv\mu^0$ and $I\equiv I^0$ together with $\alpha={\mu_b}+2\mu/3$ and $\beta=2/I$.
%
%

\end{document}